\documentclass[12pt]{article}
\usepackage{graphicx}
\usepackage{a4wide}
\begin{document}
\title{A toy model for generalised parton distributions}
\author{J.R.~Cudell\footnote{JR. Cudell@ulg.ac.be}~\footnote{Physique th\'eorique fondamentale, D\'ep. de  Physique B5a, Univ. de Li\`ege, B-4000 Li\`ege, Belgium}, F. Bissey\footnote{Institute of Fundamental Sciences, Massey Univ., P.B. 11 222, Palmerston North, New Zealand}, J.~Cugnon$^\dagger$, M.~Jaminon$^\dagger$, \\ J.P.~Lansberg$^\dagger$, and P.~Stassart$^\dagger$}
\maketitle
{\small \noindent pacs: {13.60.Hb, 13.60.Fz, 14.40.Aq, 12.38.Aw.}\\
keywords: {parton distributions, GPD, bound states, gauge invariance.\\}
}
\vskip 1cm
\centerline{\bf Abstract}
\begin{quote}
We give the results of a simple model for the diagonal and off-diagonal 
valence quark distributions of a pion. We show that structure can be implemented
in a gauge-invariant manner. This explicit model questions the validity
of the momentum sum rule, and gives an explicit counter-example to the 
Wandzura-Wilczek ansatz for twist-3 GPD's.
\end{quote}


\section{Introduction}
We review the results on structure functions
and generalised parton distributions (GPD's) that we obtained 
some time ago in \cite{us}. Our purpose is to explore the simplest 
model of a hadron in order to have an explicit representation 
and some intuition 
for GPD's. Hence we have considered the simplest hadron, {\it i.e.} the
pion, and concentrated on its valence-quark content. In order to 
further simplify the model, we took a $\pi^0$, although our twist-2
and twist-3 results remain true for $\pi^\pm$, as the diagrams involving the direct $\gamma\pi^\pm$ coupling are suppressed by powers of $Q^2$. 

\section{Diagonal structure}
\subsection{Structureless pions}
\begin{figure}[h]
\centerline{\includegraphics[width=.30\textwidth]{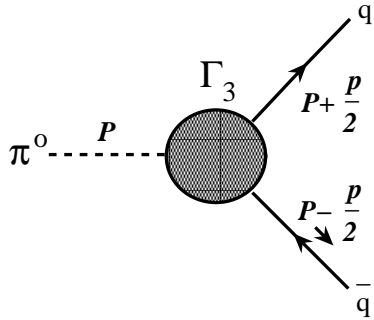}}
\caption{The lowest-order vertex $\pi^0 q\bar q$.}
\label{fig:vertex}
\end{figure}

We must first consider the diagonal structure functions. In order to
calculate them explicitly, we must make a model for the pion. 
The minimal requirement is that it is a pseudo-scalar made of a 
quark and an antiquark. Hence we must select the proper spin 
states, which is easily 
done through the use of a $\gamma_5$ vertex. The simplest assumption
is then that the pion can be treated as a point-like particle, 
coupling to quarks via an effective 3-point 
vertex (shown in Fig.\ref{fig:vertex}):
\begin{equation} \Gamma_3=i
g\gamma_5,\label{vertex}\end{equation} with $g$ a coupling constant.

\begin{figure}[b]
\centerline{\includegraphics[width=.40\textwidth]{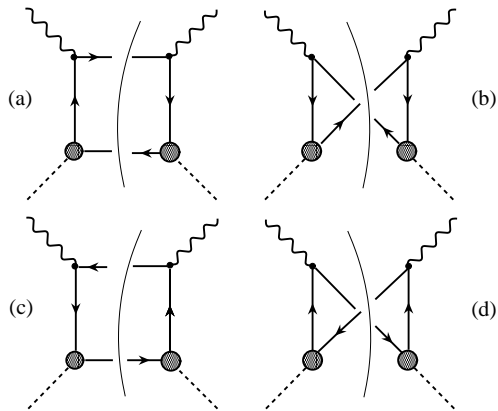}}
\caption{The four cut diagrams contributing to structure functions.}
\label{fig:cuts}
\end{figure}
The lowest-order approximation for the structure functions $F_1$
and $F_2$ then comes from the discontinuity ${\rm disc}({\cal A}_{\mu\nu})$
 of the diagrams of
Fig. \ref{fig:cuts}.
In order to make the calculation infrared finite (or at least to avoid the
poles on the vertical lines of the diagrams of Fig.~\ref{fig:cuts}), we 
assume that the 
quarks are sufficiently massive:
\begin{equation} 2m_q>m_\pi\end{equation}
(We shall take $m_q\approx 300$ MeV in the following.)
The discontinuity is then obtained by putting the intermediate states on-shell.
The answer one gets is explicitly gauge-invariant, and one may obtain the
structure functions via the usual formula:
\begin{equation}
\frac{1}{4\pi} {\rm disc}({\cal A}_{\mu\nu})
= \left(-g_{\mu
\nu}+\frac{q_{\mu}q_{\nu}}{q^2}\right) F_1 +
\left(P_{\mu}-q_{\mu}\frac{P.q}{q^2}\right)
\left(P_{\nu}-q_{\nu}\frac{P.q}{q^2}\right)(P.q) F_2, \label{W}
\end{equation}
 with $P$ the momentum of the pion and $q$ that of the photon.

In the Bjorken limit, $Q^2\rightarrow\infty$, $x=Q^2/(2~P.q)$
fixed, one obtains
\begin{equation}
F_1=\frac{5 g^2}{24 \pi^2} \left[ \log\left(\frac{(1-x) Q^2}{xM^2}\right) -
\frac{m_{\pi}^2} {M^2} x (1-x)  \right], \label{Lnu}
\end{equation}
with $M^2=m_q^2-m_{\pi}^2 x (1-x)$.
\begin{figure}
\centerline{\includegraphics[width=.60\textwidth]{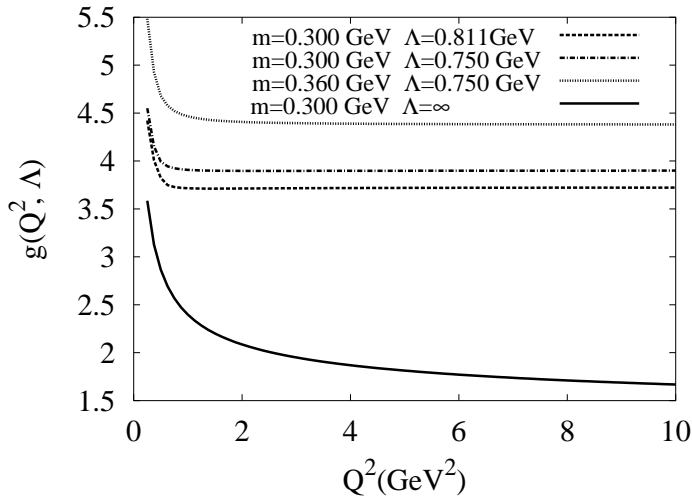}}
\caption{The coupling constant of the vertex.}
\label{fig:norma}
\end{figure}

One still needs however to determine the coupling $g$ of the vertex 
(\ref{vertex}). In principle, this can be done via the electromagnetic
form factor at $t=0$. We find it easier to use the Adler sum rule, which
should be equivalent for the leading twist:
\begin{equation}
\int_0^1 dx~~F_1(x,Q^2)={5\over 18}.
\end{equation}
It amounts to saying that our pion is made of $u\bar u$ and $d\bar d$ with
equal probability. It leads to a coupling $g$  given by the lower curve of Fig.~\ref{fig:norma}.
One obtains the
Callan-Gross relation $F_2(x,Q^2)=2x F_1(x,Q^2)$, and a prediction
for $F_1$.

However, we see that, because $F_1$ has a logarithmic growth at fixed $g$, the
normalisation condition actually makes $g$ run down as $1/\log Q^2$. This is not consistent with the definition (\ref{vertex}) of the vertex: $g$ can depend only 
on the variables entering the vertex, {\it i.e.} $P$ and $p$, and
cannot depend directly on $Q^2$.

Furthermore, although the diagrams of Fig. \ref{fig:cuts}.a and \ref{fig:cuts}.c have a probabilistic interpretation,
they are not enhanced by a power of $Q^2$ with respect the interference
graphs of Fig. \ref{fig:cuts}.b and \ref{fig:cuts}.d: we cannot define parton distributions. The reason for
this is obvious: our pion does not have a structure. 

\subsection{Structure and gauge invariance}
\begin{figure}
\centerline{\includegraphics[width=.30\textwidth]{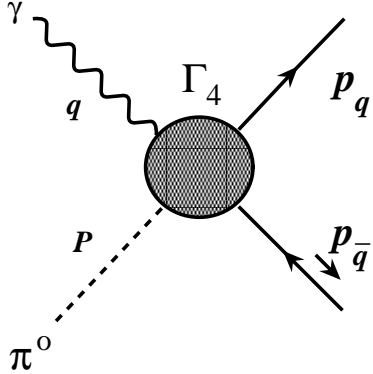}}
\caption{Representation of the 4-point vertex.}
\label{fig:4pt}
\end{figure}

Guided again by simplicity, we shall assume that it is a good approximation
to represent confinement effects by a cut on the square of the relative 
4-momentum
of the partons. The vertex of Fig. \ref{fig:vertex} now becomes a step function of 
$p^2$,
\begin{equation}
\Gamma_3(P,p)=ig(Q^2,\Lambda^2)\gamma_5\theta(|p^2|<\Lambda^2)
\label{gamma3}
\end{equation}
and we shall take $\Lambda\approx 0.8$~GeV (or $r_\pi\approx 0.25$ fm).
\begin{figure}[h]
\centerline{\includegraphics[width=.60\textwidth]{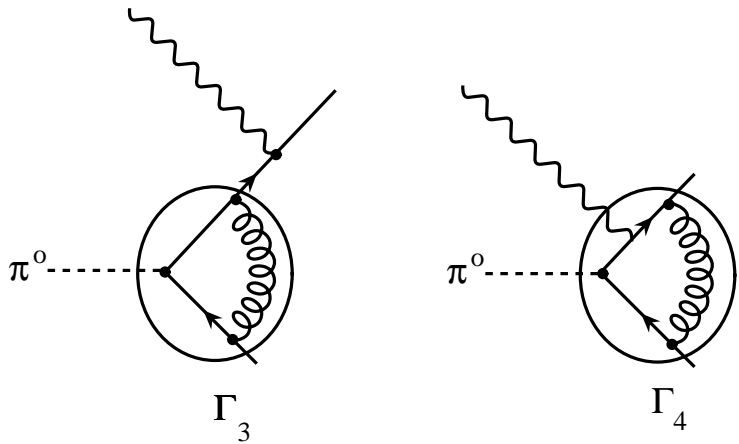}}
\caption{Illustration of the necessity of a 4-point vertex.}
\label{fig:4pts}
\end{figure}
But the introduction of structure has a dire consequence: this simple modification 
is not sufficient as it breaks gauge invariance. This can easily 
be understood
if we represent the vertex function by an exchange, as in Fig. \ref{fig:4pts}. To get a 
complete set of diagrams leading to a gauge-invariant answer, on must include
both diagrams in Fig. \ref{fig:4pts}. The first diagram is analogous to a simple modification
of the 3-point vertex, such as (\ref{gamma3}), whereas the second one
can only be included in a 4-point vertex, as in Fig.~\ref{fig:4pt}.

\begin{figure}[h]
\centerline{\includegraphics[width=.40\textwidth]{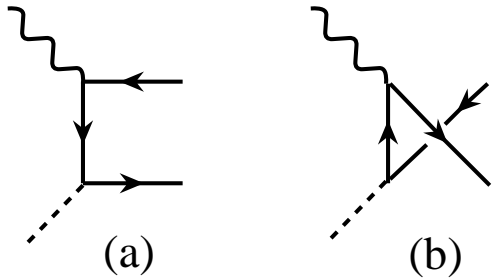}}
\caption{Two amplitudes of the cut.}
\label{fig:ampl}
\end{figure}

The latter is unknown, and we shall use a simple trick to model it.
To analyse the problem, it is enough to consider one half of the cut
amplitude, as in Fig. \ref{fig:ampl}. We can define the usual Mandelstam variables
as $\hat t=(P-p_q)^2$ and $\hat u=(P-p_{\bar q})^2$. 
The cut on the relative momentum then amounts to a cut on $t$ for
diagram \ref{fig:ampl}.a ($p^2=-m_\pi^2+2m_q^2+2t$), whereas it is a cut on $u$ for diagram \ref{fig:ampl}.b ($p^2=-m_\pi^2+2m_q^2+2u$). 
Hence both diagrams 
have different physical cuts, which gives rise to the gauge-invariance problem.
The solution is then simple: one must invent a 4-point vertex such that both
graphs are cut in the same way. Hence, we need to multiply the sum of
the two graphs of Fig. \ref{fig:ampl} by the same function and, to obtain (\ref{gamma3}),
we must choose 
\begin{equation}
F(t,u)=[\theta(|-m_\pi^2+2m_q^2+2\hat t|<\Lambda^2)+
\theta(|-m_\pi^2+2m_q^2+2\hat u|<\Lambda^2)].
\end{equation}
The first term is the 3-point vertex (\ref{gamma3}), whereas the second term
can be interpreted as a contribution from a new 4-point vertex, shown in
Fig. \ref{fig:4pt}:
\begin{equation}
\Gamma_4(p_q,p_{\bar q},P,q)=g(Q^2,\Lambda^2) \theta(|-m_\pi^2+2m_q^2+2\hat u|
<\Lambda^2) {\gamma_\mu (\gamma\cdot(P-p_{\bar q})+m_q)\gamma_5\over (P-p_{\bar q})^2-m_q^2}
\end{equation}

\begin{figure}[h]
\centerline{\includegraphics[width=.45\textwidth]{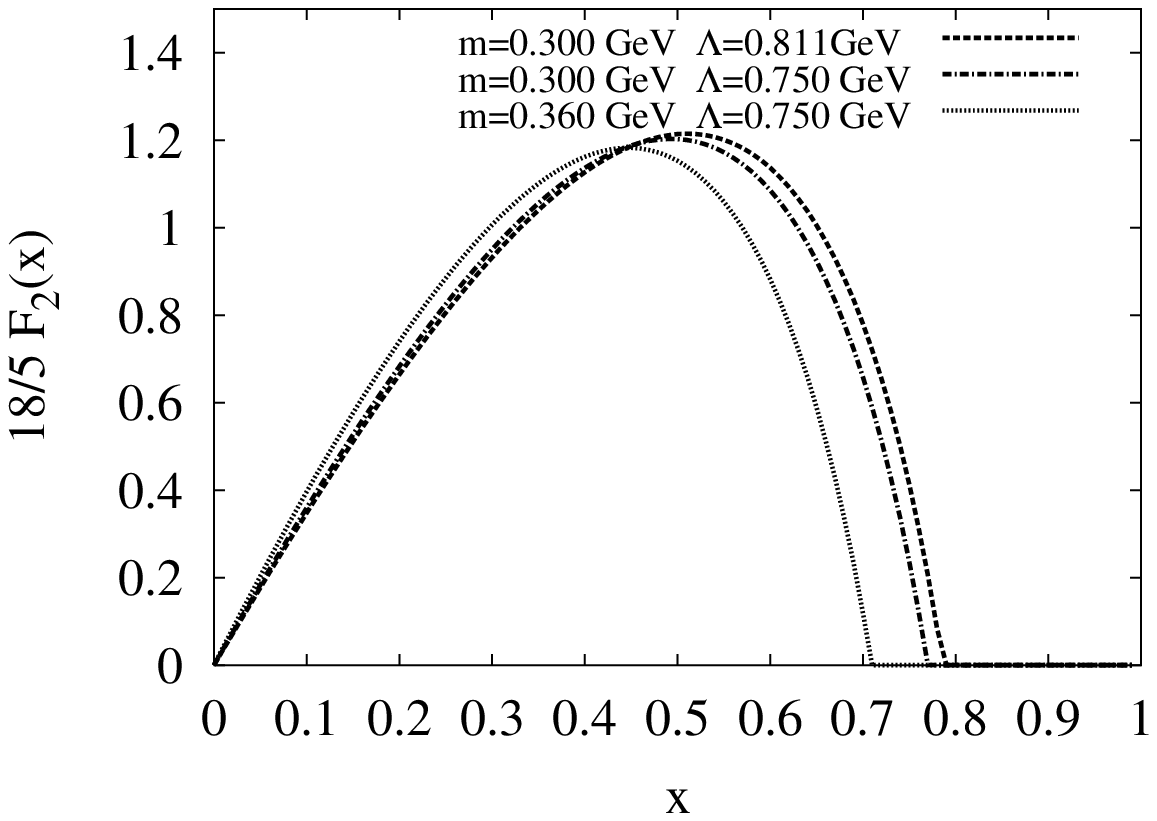}
\includegraphics[width=.45\textwidth]{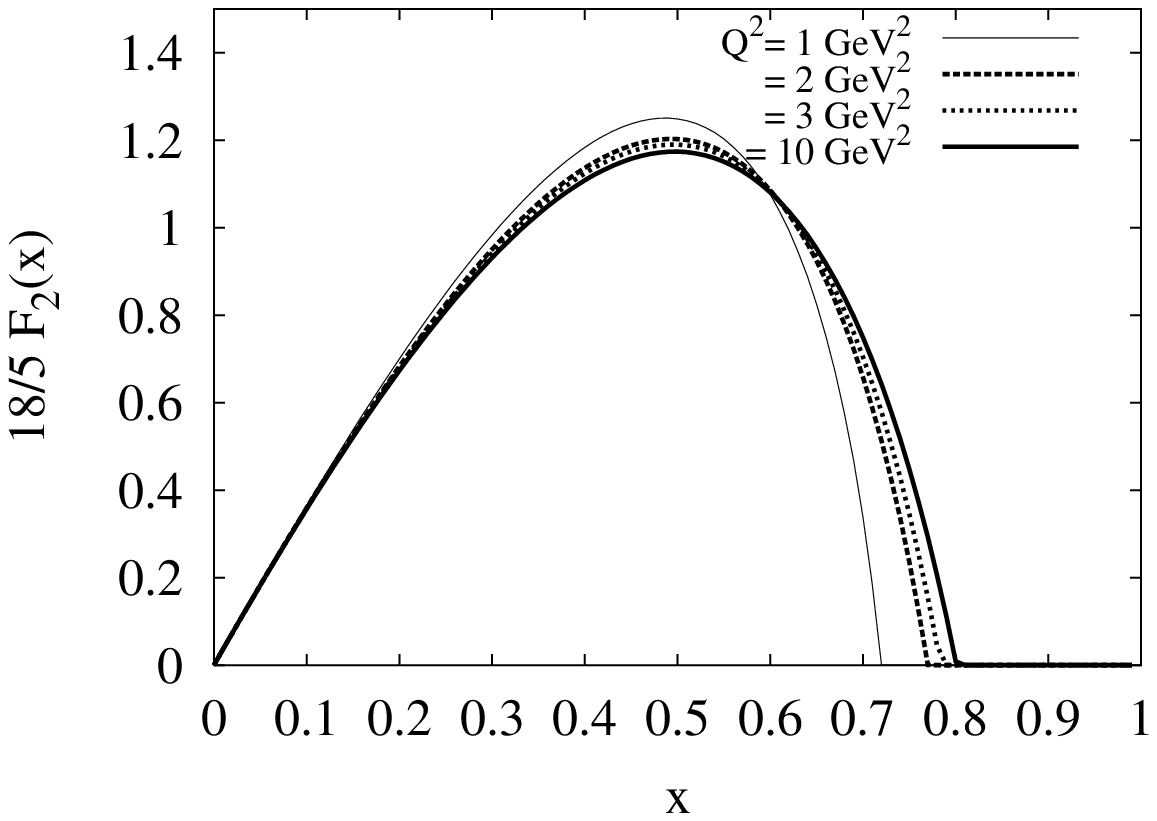}}
\caption{Results for the diagonal structure function. Left: for 
$Q^2=2$~GeV$^2$, right: $m_q=0.3$~GeV, $\Lambda=0.75$~GeV.}
\label{fig:F1}
\end{figure}

One can then re-calculate $F_1$ and $F_2$, and normalise them again.
The coupling constants are shown in Fig.~\ref{fig:norma}. We see
that $g$ can now be taken as a constant for values of $Q^2$ large enough
for the Adler sum rule to hold. 
The curves for $F_2$ are given in Fig. \ref{fig:F1}, for various choices of the cut-off $\Lambda$ and of the quark masses\footnote{Please note that the curves
are for $18/5 F_2$. The factor $18/5$ was missed in the first paper of \cite{us}.}.

One can see that 
$F_2$ is stable w.r.t. $Q^2$, so that the structure
function seems a possible candidate for the initial valence quark 
distribution in a pion $u_v(x)=\bar{u}_v(x)=d_v(x)=\bar{d}_v(x)=v(x)$, 
via the relation
\begin{equation}
F_1= \frac{5}{9}v(x).
\label{vx}
\end{equation} 

Note however that the condition 
$s>4m_q^2$ leads to a cut on the values of $x$ which are allowed:
whereas for no cut-off, $x$ can go to 1 in the limit 
$Q^2\rightarrow\infty$, it is limited to $x<1-(m_q/\Lambda)^2$ for 
finite $\Lambda$. Although the elimination of the interval close to $x=1$
is due to the fact that our cut-off is sharp, the suppression at large $x$
is reminiscent of that obtained in the covariant parton model \cite{LS}.
In this model, a vertex falling as $1/(p^2)^n$ at large $p$ 
leads to a parton distribution that behaves like $(1-x)^{(n-1)}$. 
 
\begin{figure}
\centerline{\includegraphics[width=.50\textwidth]{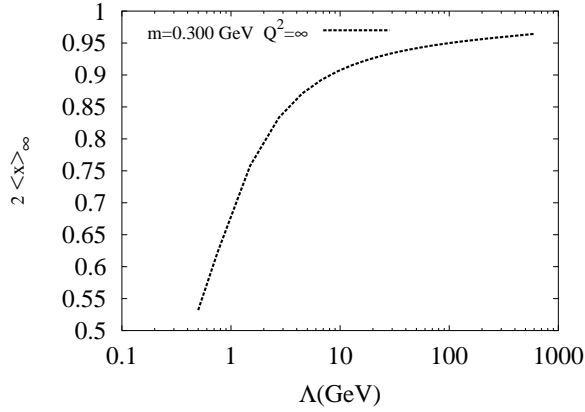}}
\caption{The average momentum carried by the quarks.}
\label{fig:xave}
\end{figure}

This leads to one of the puzzles of this model. We can consider the average
momentum carried by the quarks
\begin{equation}
\label{mf}
2\left<x\right>=4\int _{0}^{1}xv(x)dx=\frac{18}{5}\int _{0}^{1}F_{2}(x)dx.
\end{equation}
As we do not have gluons in the model, one would expect $2<x>$ to be equal to 1
as $Q^2\rightarrow \infty$. However, we find that it is in fact significantly lower, as shown in Fig. \ref{fig:xave}.
In fact, the momentum sum rule is correct either for $\Lambda=\infty$ ({\it i.e.} for structureless pions), or in the case $m_q<m_\pi/2$ (as
the infrared divergence can be re-absorbed in the normalisation). 

Because the momentum sum rule is obeyed in the structureless case, and because high momenta are cut off in our model (or in the covariant parton model), it
is obvious that $2<x>$ must be smaller than 1.  
There are two possible conclusions: the first is that the momentum sum
rule holds only for free partons, so that it is not realised in our model,
in which partons are always off-shell. But then physical quarks are always
off-shell, so one may wonder if the sum rule is true. On the other hand,
one may argue that the problem is that we did not consider cuts through the
vertices, which would lead to a gluonic component. One must then understand
why those cuts would have no effect on the sum  rule if $m_q<m_\pi/2$, whereas
they would increase it by a factor 2 for higher masses. Whatever the scenario,
the conclusion is that our calculation is reasonable for valence quarks.
 
\subsection{Off-diagonal structure}
It is easy to extend our previous calculation to the off-diagonal case. Here,
we shall again calculate the discontinuity of the amplitude, but with an
off-diagonal kinematics: the external photons have  incoming momentum $q_1$ and outgoing momentum $q_2$, whereas the pions have momenta $P_1$ and $P_2$.
We define average momenta $P=(P_1+P_2)/2$ and $q=(q_1+q_2)/2$, and the momentum
transfer $\Delta=P_2-P_1$. The Lorentz invariants of the process are $t=\Delta^2$, $Q^2=-q^2$, $x=Q^2/(2P.q)$, and $\xi=\Delta.q/(2 P.q)$.

\begin{figure}
\centerline{\includegraphics[width=.50\textwidth,clip=true]{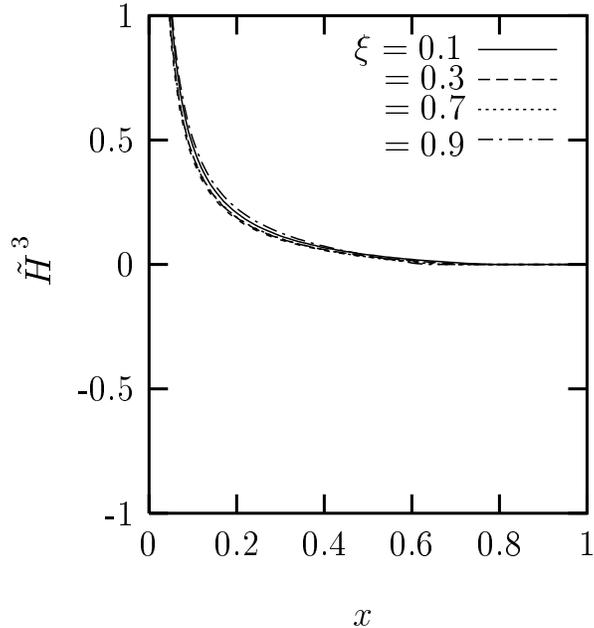}}
\caption{Our prediction for the valence component of $\tilde H^3$,
for $Q^2=10$~GeV$^2$, $t=-0.1$ GeV$^2$ and $m_\pi=0$.}
\label{fig:H3}
\end{figure}

The calculation proceeds as in the diagonal case\footnote{although we 
simplify the results by setting the pion mass to zero in the 
off-diagonal case.}: we calculate the discontinuity using 
the vertices $\Gamma_3$ and $\Gamma_4$ defined above. The answer is
again gauge invariant, and can now be decomposed into 5 independent structures
\cite{Muller}.
\begin{eqnarray}
\label{tmunudecomp}
\frac{1}{4\pi} {\rm disc}({\cal A}_{\mu\nu})
&=& - {\cal P}_{\mu\sigma} g^{\sigma\tau} {\cal P}_{\tau\nu}
F_1
+ \frac{{\cal P}_{\mu\sigma} p^\sigma p^\tau {\cal P}_{\tau\nu}}{p \cdot q}
F_2\nonumber\\
&+& \frac{{\cal P}_{\mu\sigma} (p^\sigma (\Delta^\tau-2\xi p^\tau)
+  (\Delta^\sigma-2\xi p^\sigma) p^\tau) {\cal P}_{\tau\nu}}{2 p \cdot q}
F_3\nonumber\\
&+& \frac{{\cal P}_{\mu\sigma}
(p^\sigma (\Delta^\tau-2\xi p^\tau)
-  (\Delta^\sigma-2\xi p^\sigma)p^\tau) {\cal P}_{\tau\nu}}{2 p \cdot q}
F_4\nonumber\\
&+& {\cal P}_{\mu\sigma}
(\Delta^\sigma-2\xi p^\sigma) (\Delta^\tau-2\xi p^\tau) {\cal P}_{\tau\nu}
F_5    ,
\end{eqnarray}
where we have used the projector ${\cal P}_{\mu\nu} = g_{\mu\nu} - \frac{q_{2 \mu} q_{1 \nu}} {q_1 \cdot q_2}$.
For neutral pions, the structure functions $F_i$ that parameterize
the discontinuity can be directly related to the GPD's $H$, $H^3$
and $\tilde H^3$ \cite{Muller} to twist-3 accuracy:
\begin{eqnarray}
{1\over 2\pi}F_1&=&{H}, \label{eq:FH1}   \\
{1\over 2\pi}F_2&=&2x{H}+{\cal O}({1}/{Q^2}), \label{eq:FH2}  \\
{1\over 2\pi}F_3&=&\frac{2x}{x^2-\xi^2}\left({H}^3x^2+\tilde{H}^3\xi x-{H}\xi\right)+{\cal O}({1}/{Q^2}), \label{eq:FH3}    \\
{1\over 2\pi}F_4&=&\frac{2x}{x^2-\xi^2}\left({H}^3\xi x+\tilde{H}^3x^2-{H}x\right)+{\cal O}({1}/{Q^2}), \label{eq:FH4}\\
{1\over 2\pi}F_5&=& {\cal O}({1}/{Q^2}).\label{eq:FH5}
\end{eqnarray}
Our calculation obeys Eqs. (\ref{eq:FH2}) and (\ref{eq:FH5}), and 
leads to definite predictions for $H$, $H^3$ and $\tilde H^3$. 
Before giving the explicit results, let us mention that we find 
explicit relations linking the twist-3 GPD's to $H$:
\begin{eqnarray}
{H}^3&=&\frac{(x-1)\xi}{x(\xi^2-1)}H+{\cal O}({1}/{Q^2})\label{H3}\\
\tilde{H}^3&=&\frac{H^3}{\xi}+{\cal O}({1}/{Q^2})\label{H3t}
\end{eqnarray} 
Note that the polynomiality of the Mellin moments of $H$, $H^3$ 
and $\tilde H^3$, together with Eqs.~(\ref{H3}) and (\ref{H3t}), 
implies that
$H$ must be a polynomial $P_H(\xi)$ multiplying $\xi^2-1$. We show in Fig. \ref{fig:H3} our results for $\tilde H^3$. The fact that
it is almost independent of
$\xi$ shows that $P_H$ is very close to a constant.

Let us point out that the relations (\ref{H3}, \ref{H3t}) are
an explicit counter-example to the Wandzura-Wilczek ansatz \cite{Muller,WW}. Not
only are they numerically different, but our calculated $H^3$ and $\tilde H^3$
do not suffer from discontinuities at $x=\xi$, contrarily to what the 
Wandzura-Wilczek ansatz predicts.

\begin{figure}[h]
\centerline{\includegraphics[width=.5\textwidth,clip=true]{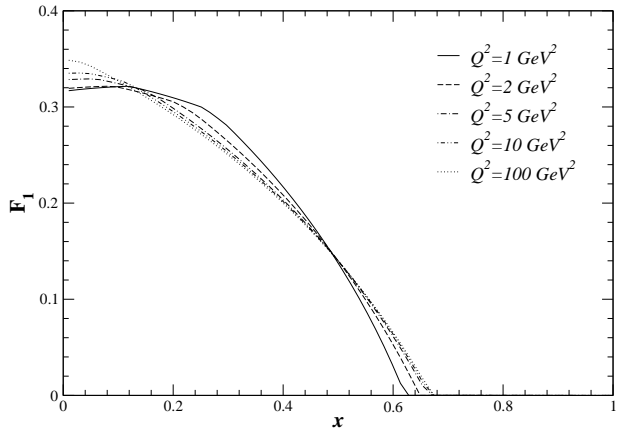}}
\caption{Our prediction for the valence component of $H$,
in DVCS, as a function of $Q^2$, for $t=-0.1$~GeV$^2$,  $\Lambda=0.75$~GeV and $m_\pi=0$.}
\label{fig:Q2evol}
\end{figure}

We can finally turn to our predictions for $H$. Here we consider two regimes:
deeply virtual Compton scattering ($\xi=-x$), and elastic scattering ($\xi=0$).
First of all, we show in Fig. \ref{fig:Q2evol} that our ansatz is stable w.r.t.
$Q^2$. It can presumably be used as the initial parton distribution, as
in the diagonal case.\vglue 0.5cm

\begin{figure}[h]
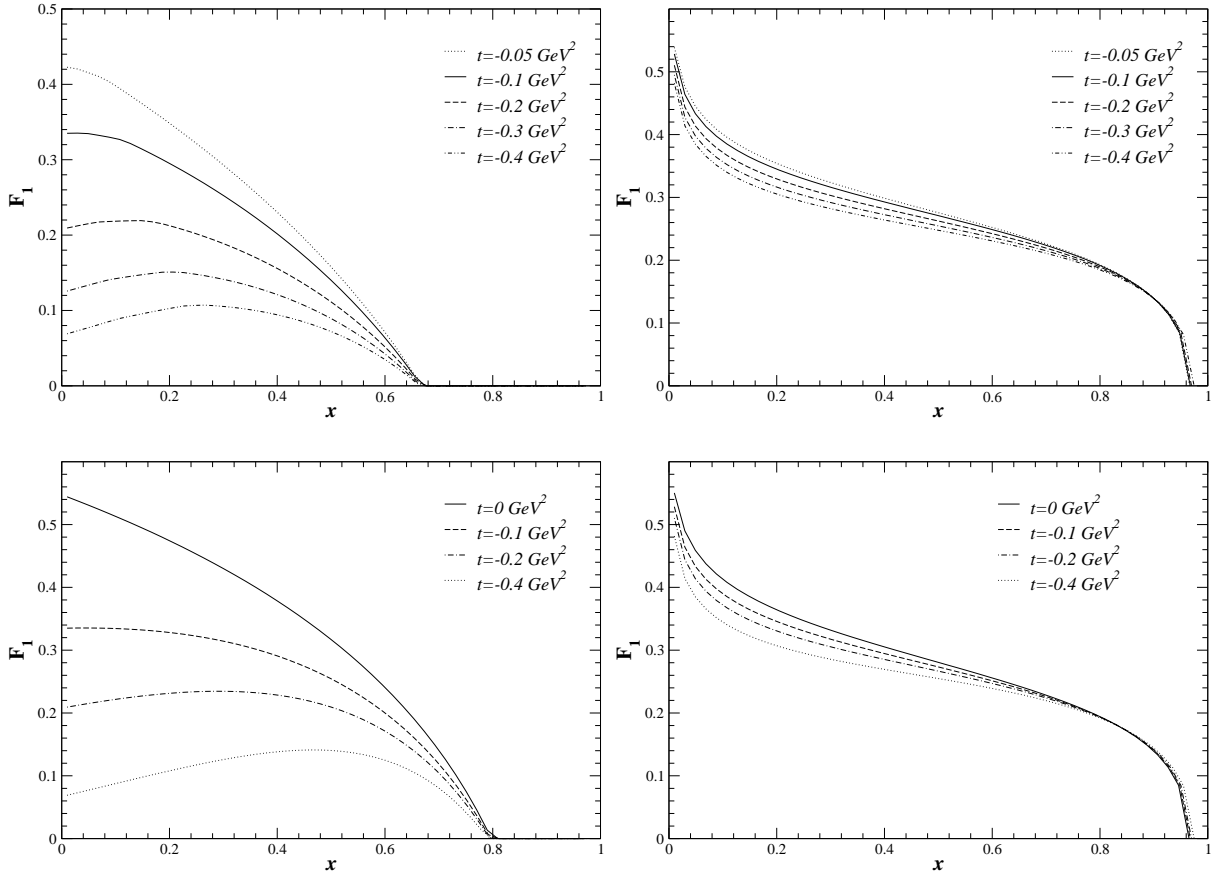

\vbox{
\centering
{~\\ \includegraphics[width=.49\textwidth,clip=true]{F1dvcs-q2_10-lam_0_75.eps}}
{\includegraphics[width=.49\textwidth,clip=true]{F1dvcs-nco-q2_10.eps}}\\~\\
{\includegraphics[width=.49\textwidth,clip=true]{F1-xi_0-q2_10-lam_0_75.eps}}
{\includegraphics[width=.49\textwidth,clip=true]{F1-nco-xi_0-q2_10.eps}}}
\caption{Our prediction for the valence component of $H$,
in DVCS (top) and in elastic scattering (bottom), 
as a function of $t$, for $Q^2=$ 10~GeV$^2$ and $m_\pi=0$. The left
graphs are for $\Lambda=0.75$~GeV and the right ones for structureless pions.}
\label{fig:DVCS}
\end{figure}

We can also examine the role of hadronic 
structure by comparing our prediction
at $\Lambda=0.75$~GeV with that for a structureless pion ($\Lambda=\infty$).
We do this both in the DVCS and in the elastic cases in Fig. \ref{fig:DVCS}. 
We see that the structure of the pion makes an enormous difference. 
The cut-off in $x$ is in fact smaller than in the diagonal case, and the
novelty is a rather large dependence on $t$. Hence we confirm that DVCS, and
GPD's in general, will give us new information about hadronic structure.\\~\\

\section{Conclusions}
We have built a very simple model for the pion, which goes beyond the
spectator quark model. It implements all the symmetries of the problem,
in particular gauge invariance. This toy model has enabled us to study
explicitly initial valence quark (and antiquark) distributions, both in
the diagonal and in the off-diagonal cases. We find that structure leads to
corrections of order $m_q/\Lambda$ to the momentum sum rule in the diagonal 
case. Such effects might be attributed to the contributions of cuts in the
vertices. Using this model in the off-diagonal case, we have shown that
the Wandzura-Wilczek ansatz, which can be used to relate twist-3 GPD's 
to twist-2, is likely to be wrong. We in fact obtain new relations between
the GPD's. We have also shown that binding effects lead to a rich structure
for GPD's, which is not present in the case of point couplings.

\section*{Acknowledgments}
We thank M.V. Polyakov and P.V. Landshoff for discussions. J.P.L. 
is a research fellow of the Institut Interuniversitaire des Sciences
Nucl\'eaires.


\begin{thebibliography}{9}

\bibitem{us} 
F.~Bissey, J.~R.~Cudell, J.~Cugnon, J.~P.~Lansberg and P.~Stassart,
Phys.\ Lett.\ B {\bf 587} (2004) 189
[arXiv:hep-ph/0310184];
F.~Bissey, J.~R.~Cudell, J.~Cugnon, M.~Jaminon, J.~P.~Lansberg and P.~Stassart,
Phys.\ Lett.\ B {\bf 547} (2002) 210
[arXiv:hep-ph/0207107].
\bibitem{LS}  P.~V.~Landshoff, J.~C.~Polkinghorne and R.~D.~Short,
  Nucl.\ Phys.\ B {\bf 28} (1971) 225;
  P.~V.~Landshoff and D.~M.~Scott,
  Nucl.\ Phys.\ B {\bf 131} (1977) 172.
\bibitem{Muller}A.~V.~Belitsky, D.~M\"uller, A.~Kirchner and A.~Sch\"afer,
Phys.\ Rev.\ D {\bf 64} (2001) 116002
[arXiv:hep-ph/0011314].
\bibitem{WW}S.~Wandzura and F.~Wilczek,
Phys.\ Lett.\ B {\bf 72} (1977) 195. 
\end{thebibliography}
\end{document}